\documentclass[twocolumn,pre,showpacs,preprintnumbers,amsmath,amssymb]{revtex4}

\usepackage{graphicx}
\usepackage{dcolumn}
\usepackage{bm}

\begin{document}
\title{\textbf{\textrm{Multi-partite entanglement and quantum phase transition
in the one-, two-, and three-dimensional transverse field Ising model}}}

\author{Afshin Montakhab}
\email{montakhab@shirazu.ac.ir}

\author{Ali Asadian}

\affiliation{Department of Physics, College of Sciences, Shiraz University, Shiraz 71454, Iran}

\date{\today}
\begin{abstract}
In this paper we consider the quantum phase transition in the Ising model in the
presence of a transverse field in one, two and three dimensions from a multi-partite
entanglement point of view.  Using \emph{exact} numerical solutions, we
are able to study such systems up
to 25 qubits. The Meyer-Wallach measure of global entanglement
is used to  study the critical behavior of this model.
The transition we consider is between a symmetric GHZ-like state to a paramagnetic
product-state.  We find that global entanglement serves
as a good indicator of quantum phase transition with interesting scaling behavior.
We use finite-size scaling to extract the critical point as well as some critical exponents
for the one and two dimensional models. Our results indicate that such multi-partite measure of global
entanglement shows universal features regardless of dimension $d$.  Our results also provides
evidence that multi-partite entanglement is better suited  for the study of quantum phase transitions
than the much studied bi-partite measures.
\end{abstract}

\maketitle

\section{Introduction}

There has been much work on entanglement in the past twenty years \cite{Peres, Werner, Plenio07, Ter02, Van07, Audbook, Bruss02, Hord07}.
Entanglement is a purely quantum phenomenon with no classical counterpart.  It is thought to hold the key to a deeper
understanding of the theoretical aspects of quantum mechanics.  From a more practical aspect, entanglement is the key ingredient in many information processing applications including quantum computation and quantum cryptography \cite{NC2000}.  On the other
hand, there has been much cross-fertilization in the fields of condensed-matter physics
and quantum information theory in recent years \cite{Amico, Ost02, Osb02, Arensen, Latorre, Vidal, Duer, Guehne, Wang}.  Here, many traditional
condensed-matter systems including fermionic and bosonic gases, and in particular lattice
spin models have been investigated in the light of new developments in quantum information
theory and entanglement in particular \cite{Amico}.  It has been found that entanglement plays a crucial
role in the low-temperature physics of many of these systems, particularly in their
ground (zero-temperature) state \cite{Ost02, Osb02, Vidal,LL06}.  A very fruitful avenue
along these lines has been the relation of entanglement and phase transitions in general, and
quantum (ground state) phase transitions in particular. This is a bit surprising since entanglement was originally
thought to be somewhat fragile and thus easily destroyed by fluctuations.

In a quantum phase transition (QPT) \cite{Sachdev}, a thermodynamic system described by a Hamiltonian
$H(\lambda)$ changes its macroscopic phase at the critical value of the control
parameter $ \lambda_c$.  In recent studies of many thermodynamic systems exhibiting QPT,
in particular quantum spin models\cite{Osb02, Ost02, Vidal,LL06}, it has become clear that the onset of transition is accompanied
by a marked change in the entanglement.  Depending on the model, entanglement could peak, or show
discontinuous behavior, or show diverging derivatives with scaling behavior at the critical point\cite{Amico,GTL07}.
What is less clear is the exact role (or the general mechanism) through which entanglement and
QPT are related.  In such studies, various quantitatively different measures of entanglement have been used.
Therefore, for example, one would like to know if there are universal features in the entanglement of various spin
model exhibiting QPT?

Another important feature which is emerging out of recent studies of condensed-matter systems
from quantum information perspectives is the need for multi-partite measures of entanglement
\cite{OlivPRL, Wei05, Huang10, Anfossi05, Oliv06} .
This, by the way, is an example of cross-fertilization referred to earlier.  Since the root of
quantum theory\cite{EPR} is originally in bi-partite systems like Bell states, it has been natural to
study macroscopic systems using bi-partite measures such as the von Neumann entropy or concurrence.
In fact, with very few exceptions, the general body of the current literature has used such
bi-partite measures to study many particle systems.  Although this has been so because of a
matter of tradition and/or convenience, there is increasing evidence that such measures
are generally inadequate to study QPT in condensed-matter systems\cite{yang05, Amico06}.
After all, it is natural to use multi-partite
entanglement if one is to study the role of entanglement in multi-partite (many-particle) systems,
as important types of entanglement in such systems (e.g. various $n$-tangles) may not be captured by a bi-partite measure, but would be
included in an (ideal) multi-partite measure.  Additionally, some multi-partite measures (as will be discussed
in Sec. III) have thermodynamic properties (e.g. extensivity) which make them more suitable for studies
of such thermodynamic phenomenon as QPT.
Another equally important shortcoming is that most such studies have been carried out for one dimensional (1d) models.  Although this
is perhaps because of computational difficulties, it is certainly not well-justified.  As is well-known,
spatial dimension (d) plays an important role in the physics of thermodynamic systems, phase transitions
in particular\cite{gold}.

Here, we propose to study QPT in prototypical transverse field quantum Ising model using the Meyer and Wallach \cite{MW}
measure of global entanglement in one, two, as well as three dimensions.
Such global entanglement measure seems to be well-suited for studies of
many particle systems \cite{OlivPRL, Mont}. Since analytic results are usually difficult to come up with,
numerical results with finite-size systems are typically the way to proceed.  However, solving quantum
lattice spin systems numerically is also computationally expensive as only a few qubits (spins) can
be solved exactly and approximation techniques have limited success in one dimension and are more
limited in higher dimensions \cite{langari}.  Very recently, however, such systems
have been studied using efficient numerics\cite{VidalPRL101}.

In this article, we solve the transverse field quantum Ising model numerically (\emph{exact})
for up to 25 qubits in one, two and three dimensions. Our main result is that global
entanglement is a well-suited measure to study QPT with some universal
features in any dimension.  We show that global entanglement has interesting scaling
properties near the critical point. Using finite-size scaling arguments,
we extract critical points as well as some critical exponents for
the 1d and 2d models consistent with previous studies.
Due to system-size limits, we are only able to study the smallest 3d system and thus cannot perform finite-size
studies. However, the general shape of global entanglement in the 3d model (Fig.~7) indicates that our
1d and 2d results easily generalize to 3d systems.
More importantly, our results provide a general framework for computation of an accessible
measure of entanglement and its relevance to QPT's in many-particle thermodynamic
systems.

This paper is structured as follows: in Section II, we discuss the multi-dimensional quantum Ising model in
the presence of a transverse field and its ground state properties relevant to our study here.  In Section
III, we discuss some key concepts regarding the Meyer-Wallach measure of global entanglement, while our
main results are presented in Section IV.  Our concluding remarks, including suggestions for further work
is presented in Section V.

\section{Transverse field Ising model}

The system under consideration here is the ferromagnetic Ising model in a
transverse field given by the Hamiltonian:
\begin{equation}
 H=-\lambda\sum_{<ij>} \sigma_{i}^{x}
 \sigma_{j}^{x}-\sum_{i=1}^{N}\sigma_{i}^{z},
\end{equation}
where the $\lambda=\frac{J}{B}$ where $J$ is the ferromagnetic coupling constant,
$B$ is the magnetic field, $N$ is the total number of spins (qubits) and
$\sigma_{i}^{x}$ and $\sigma_{i}^{z}$ are the Pauli spin
matrices in $x$ and $z$ direction at the site $i$. $<ij>$ means site $i$ and
$j$ being nearest neighbors on a regular d-dimensional lattice. We use
periodic boundary conditions. This model has been
extensively studied in 1d, but less is known about its properties
in 2d and 3d.  Relevant to our study here is the QPT this model exhibits
regardless of dimension.  At zero field this model exhibits ferromagnetic
behavior with net magnetization in the $x$ direction while in the large field limit,
it exhibits a paramagnetic behavior where all spins point in the field
direction $z$. The transition between these two phases occurs at the critical
value of $\lambda=\lambda_c$, in the thermodynamic limit.  It is well known that the ground
state is a product state in both these limits \cite{Jordan}.
In the first limit, the ground state is two-fold degenerate, one being a product
state of spins pointing in the positive $x$ direction $|+\rangle=|x;0\rangle_{1}|x;0\rangle_{2}...|x;0\rangle_{N}$
and the other is $|-\rangle=|x;1\rangle_{1}|x;1\rangle_{2}...|x;1\rangle_{N}$ which is the global phase flip of the
first one. In the second limit, the ground state is a product state of spins
pointing in the positive $z$ direction $|0\rangle$.
Both limits of the ground state are product states
which are disentangled but there is another possibility for the
ground state in the first limit which arises from linearity of the
Schr\"{o}dinger equation. As the
$|+\rangle$ and $|-\rangle$ are
solutions for the ground state in this limit, the superposition of
these degenerate states is also another acceptable solution for
the ground state when the applied field ($B$) tends to zero. This possibility
is a GHZ-like state which has genuine multi-qubit entanglement\cite{GHZ},
\begin{equation}
|GHZ\rangle_N=\frac{1}{\sqrt{2}}(|+\rangle+|-\rangle)
\end{equation}

The possibility of GHZ-like ground state is fascinating from a fundamental theoretical point of view as it represents a coherent superposition of two
macroscopically distinct states, and hence, is often called a cat-state.
It is proven in Ref.~\cite{Buzek} that in this
limit the ground state of the Ising model is locally unitary equivalent
to an N-partite GHZ state, and Ref.~\cite{Bayat} discusses how to prepare an Ising chain in a GHZ state using a single global control field.
It is important to note that such a ground state would show zero net magnetization, i.e. $<M_x>=0$.  Therefore, such
quantity could not be used as an order parameter to signal the phase transition under consideration here.  We also note
that such ground states have recently attracted attention from a symmetry-breaking point of view \cite{sym}.
Here, we study the multi-partite entanglement properties of such a ground state and its subsequent transition to a (paramagnetic) product-state as
a function of $\lambda$.

\section{Global entanglement}

Global entanglement, defined by the Meyer-Wallach entanglement measure
of pure-state \cite{MW}, and henceforth denoted by $E_{gl}$,  is a monotone\cite{Brennen}, and a very useful
measure of multi-partite entanglement. As we will show briefly, $E_{gl}$ is a measure
of total non-local information per particle in a general
multi-partite system. Therefore, $E_{gl}$ gives an intuitive meaning to
multi-partite entanglement as well as being an experimentally
accessible measure \cite{Brennen, Love, Boixo}.

Finite amount of information can be attributed to N-qubit pure
state which is N bit of information according to Brukner-Zeilinger
operationally invariant information measure \cite{Brukner}. This information
can be distributed in local as well as non-local form, which is
associated with entanglement \cite{Hord98}. This information has a
complimentary relation:
\begin{equation}
 I_{total}=I_{local}+I_{non-local}.
\end{equation}
The total information is conserved unless transferred to
environment through decoherence. The amount of information in
local form is $I_{local}=\sum_{i=1}^{N}I_{i}$ where,
$I_{i}=2Tr\rho_{i}^{2}-1$ is the operationally invariant
information measure of a qubit \cite{Brukner}, and $\rho_{i}$ is single
particle reduced density matrix obtained by tracing over the other
particles' degrees of freedom.  Therefore, according to Eq. (3)
$I_{non-local}=\sum_{i=1}^{N}2(1-Tr\rho_{i}^{2})$ which is
distributed in different kinds of quantum correlations, the
tangles, among the system,
\begin{equation}
 I_{non-local}=2\sum_{i_{1}<i_{2}}\tau_{i_{1}i_{2}}+...+N\sum_{i_{1}<...<i_{N}}\tau_{i_{1}...i_{N}},
\end{equation}
where the first term is referred to as 2-tangle, the next being
3-tangle and the last term the N-tangle of the system. One can view
these tangles as different types of non-local information distribution.
Therefore, since $E_{gl}$ is the sum of single particle linear entropies
per unit particle in a multi-partite system \cite{MW}, it can be written as:
\begin{eqnarray}
 E_{gl}=\frac{1}{N}[2\sum_{i_{1}<i_{2}}\tau_{i_{1}i_{2}}+...+N\sum_{i_{1}<...<i_{N}}\tau_{i_{1}...i_{N}}].
\end{eqnarray}
Therefore, $E_{gl}$ is the average of tangles per particles
($\frac{\langle\tau\rangle}{N}$), without giving detailed knowledge
of tangle distribution among the individual particles. This is much
like the average energy per particle in an interacting many-particle
system. The Meyer-Wallach measure was
originally introduced as a multi-partite entanglement to
distinguish it from bi-partite entanglement measures like entropy of
entanglement.  But, as shown above, $E_{gl}$ is an
average quantity and therefore cannot distinguish between
entangled states  which have equal
$\langle\tau\rangle$ yet different distribution of tangles, like
$|GHZ\rangle_{N}$ and $|EPR\rangle^{\otimes \frac{N}{2}}$ .
However, $E_{gl}$ can distinguish between GHZ and W states since
they have different values of $\langle\tau\rangle$, so $E_{gl}$,
like a thermodynamical variable, determines the
general amount of a property in a quantum system without giving a
detailed knowledge of its sharing among the constituents. One
expects this property of global entanglement to play an
important role in studying macroscopic properties of multi-partite
quantum systems \cite{OlivPRL,Mont}.

To obtain $E_{gl}$ in these systems, we have to calculate the
single particle reduced density matrix, $\rho_{i}$. Since we use
periodic boundary conditions, the reduced density matrix is the
same for all particles. So $E_{gl}$ reduced to linear entropy of a
single particle density matrix, $\rho_{i}$:
\begin{eqnarray}
E_{gl}=2(1-Tr\rho_{i}^{2})
\end{eqnarray}

\section{Results}

Using Eq.~(6), we can therefore easily calculate $E_{gl}$ exactly for
any dimension d, up to the limitations set by computational limits of
our numerics.

We start by showing our results for the 1d model. Figure 1 shows
$E_{gl}$ vs $\lambda$ and the inset shows its derivative for various
system sizes up to $N=24$.  The general behavior shown here is that
of $E_{gl}$ increasing slowly from its zero value at $\lambda =0$ with
a sharp transition to its large $\lambda$ value of $1$ around $\lambda_c =1$.
The critical point is better seen in the derivative (inset) which peaks
at the maximal value $\lambda_m(N)$.  As the system size increases,
the peak of the derivative sharpens and moves closer to the critical
point $\lambda_c=1$.

\begin{figure}[h]
\includegraphics[width=\columnwidth]{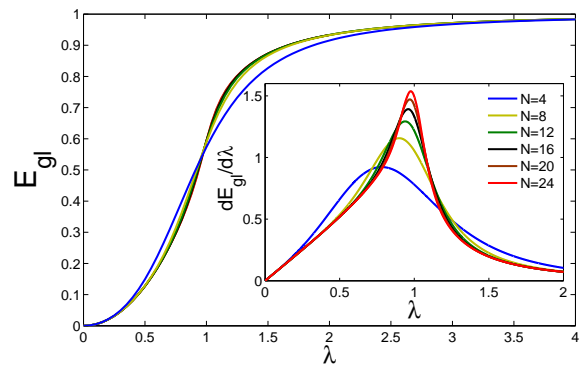}
\caption{(Color online) Global entanglement as a function of $\lambda$ for the 1d transverse
Ising model.  The inset shows the derivative and the system sizes used. Increasing $N$ sharpens
the peak and moves it closer to the critical point.}\label{3}
\end{figure}

\begin{figure}[h]
\includegraphics[width=\columnwidth]{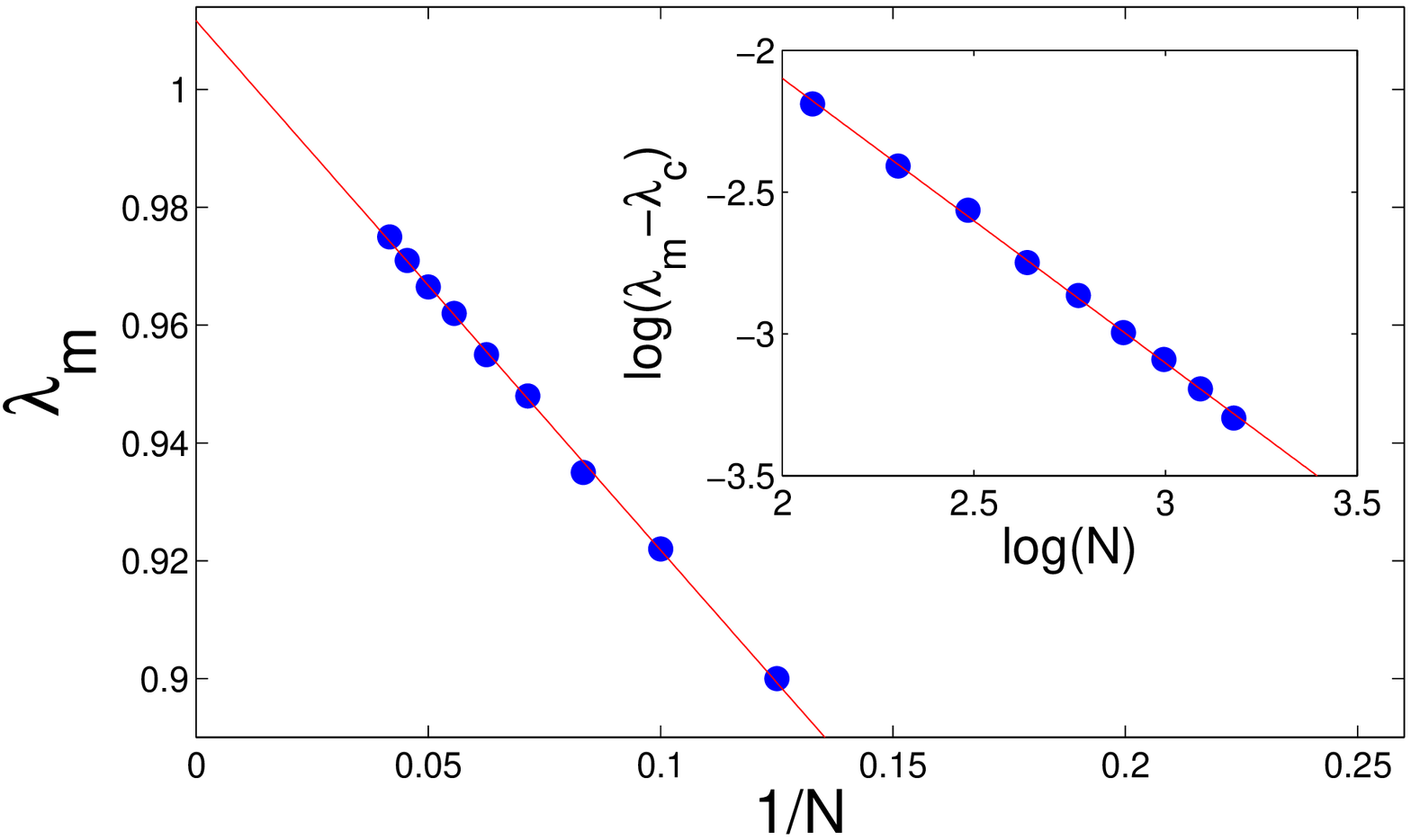}
\caption{(Color online) Convergence of $\lambda_m$ to the critical point as $N\rightarrow\infty$, for the 1d transverse Ising model.
The y-intercept is $1.01$.  The inset shows the relation $|\lambda_c - \lambda_m| \sim N^{-1.00}$.}\label{3}
\end{figure}

\begin{figure}[h]
\includegraphics[width=\columnwidth]{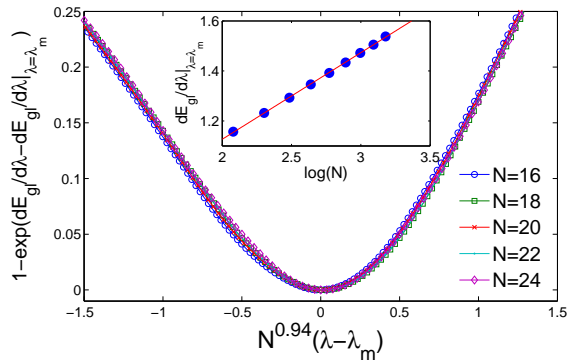}
\caption{(Color online) Finite-size scaling of global entanglement for the 1d transverse Ising model.
The inset shows the logarithmic divergence of the value of the derivative at the
maximal point $\lambda_m$.}\label{3}
\end{figure}

The extrapolation to the infinite system size along with the convergence to the
critical point (inset) is shown in Figure 2.
As one can see, $\lambda_m(\infty) =1.01$, very close to the
well-known result of $\lambda_c=1$, showing that $E_{gl}$
is a good indicator of the critical behavior of this model. The inset shows that
the convergence to the critical point is in accordance with $|\lambda_c - \lambda_m| \sim N^{-\alpha}$
with exponent $\alpha=1.00$.

We next examine the scaling behavior of $E_{gl}$ near the critical point.  According to scaling
ansatz \cite{barber}, we have $dE_{gl}/d\lambda \sim Q(N^{\frac{1}{\nu}}(\lambda-\lambda_m))$ where $\nu$
is the correlation length critical exponent, and $Q(x)\sim \ln(x)$ is generally assumed.
As is seen in Figure 3, an acceptable collapse occurs for various N's using
the scaling ansatz with the critical exponent $\nu=1.06$ in line with
previous studies \cite{Ost02}, and close to the exact result of
$\nu=1$.  The inset shows the logarithmic divergence of
the maximum (peak) of the derivative of $E_{gl}$. Hence, the general shape of $E_{gl}$, the logarithmic
divergence of its derivative at the critical point, along with its consistency with finite-size
scaling ansatz provides strong evidence for the well-suitedness of such measure for the 1d Ising model.
The question now is, if such features also hold for higher dimensional models?

We next turn to the 2d model. Using periodic boundary conditions we have
been able to study such model for up to $L^2=5^2=25=N$ qubits.
Figures 4, 5 and 6 show similar results as that of Figures 1, 2
and 3.  We note the following:  The general shape of the $E_{gl}$ still remains (Figure 4),
with a (logarithmic) divergence of the derivative at the critical point (insets of Figure 4 and 6).  The critical point
is now identified as $\lambda_m(\infty)=0.329$(Figure 5), consistent with recent studies using
infinite projected entangled-pair state ($\lambda_c=0.3268$)\cite{VidalPRL101}, as well as quantum Monte Carlo
simulations ($\lambda_c=0.3285$)\cite{QMC}.  Interestingly, our simple method obtains a more acceptable result
than the recent similar multipartite entanglement study based on matrix and tensor product states which obtained $\lambda_c=0.308$\cite{Huang10}.
We note that the convergence to the critical point (inset of Figure 5) is in accordance with $|\lambda_c - \lambda_m| \sim L^{-\alpha}$
with exponent $\alpha=1.00$ being exactly the same as the 1d case.  The difference here is that this convergence occurs from
above the critical point as opposed to the 1d case.  The finite-size scaling ansatz is also valid (Figure 6), giving the
correlation length exponent $\nu=0.51$.  The inset of Figure 6 shows the
logarithmic divergence of the derivative at the critical point.

In Figure 7, we show our result for the 3d version of this
system for the only system size we are able to study.  The general behavior of
$E_{gl}$ seen in the 1d and 2d model is clearly seen here for the 3d case as well.
While we are not able to perform scaling analysis similar to the 1d and 2d models,
it seems reasonable to assume that the same general behavior carries over to the 3d model.
We note that $\lambda_m(L=2)= 0.26$ here, which would understandably be different from
the infinite-size limit, but in the right ball-park of $\lambda_c\approx0.2$\cite{QMC}.

\begin{figure}[h]
\includegraphics[width=\columnwidth]{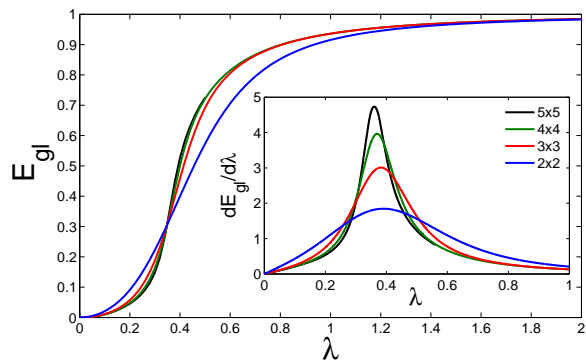}
\caption{(Color online) Global entanglement as a function of $\lambda$ for the 2d transverse Ising model.
The inset shows the derivative and the system sizes used. Increasing system sizes sharpens the peak
and moves it towards the critical point.}\label{3}
\end{figure}

\begin{figure}[h]
\includegraphics[width=\columnwidth]{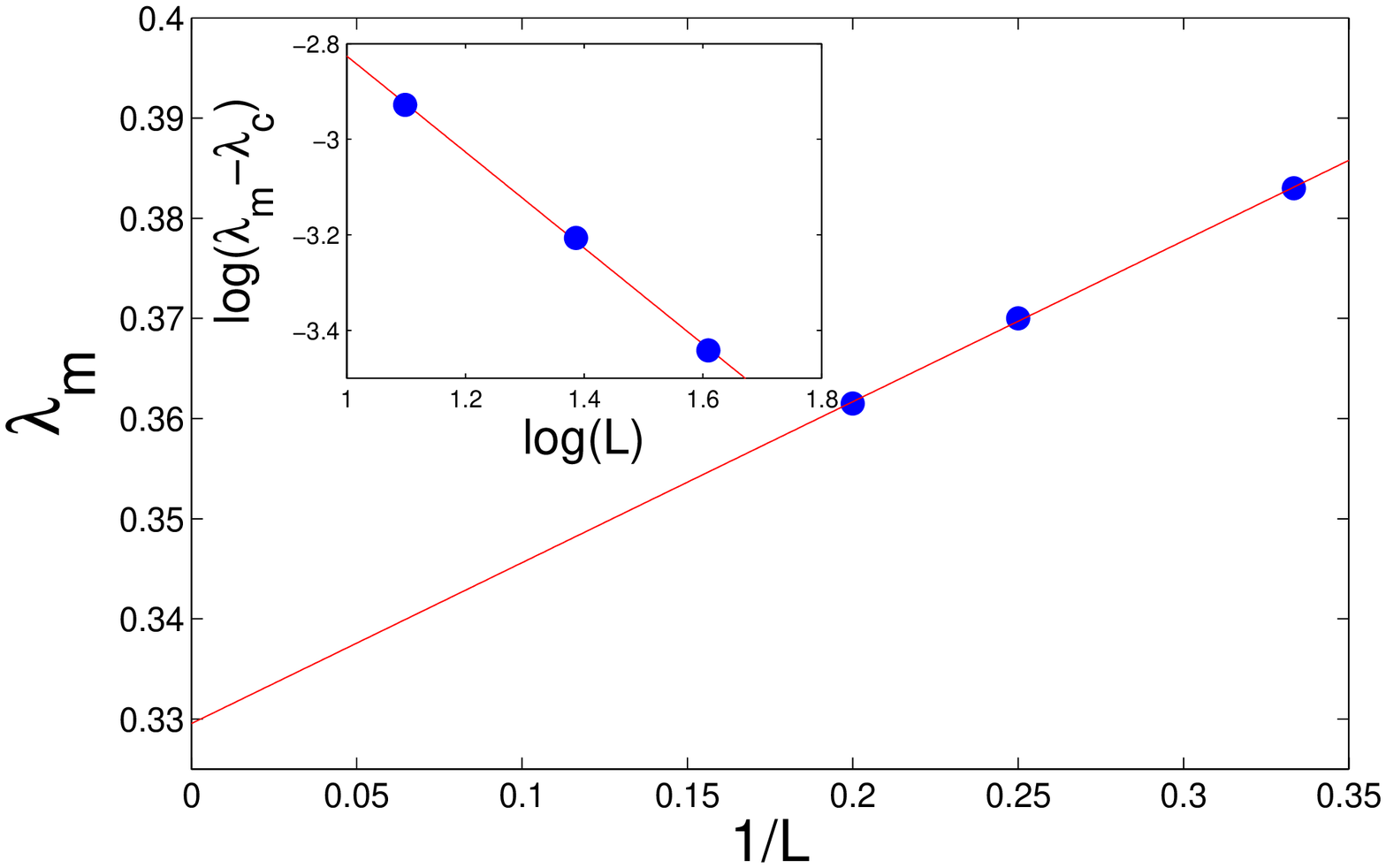}
\caption{(Color online) Convergence of $\lambda_m$ to the critical point as $L\rightarrow\infty$, for the 2d transverse Ising model.
The y-intercept is $0.329$.  The inset shows the relation $|\lambda_c - \lambda_m| \sim L^{-1.00}$.}\label{3}
\end{figure}

\begin{figure}[h]
\includegraphics[width=\columnwidth]{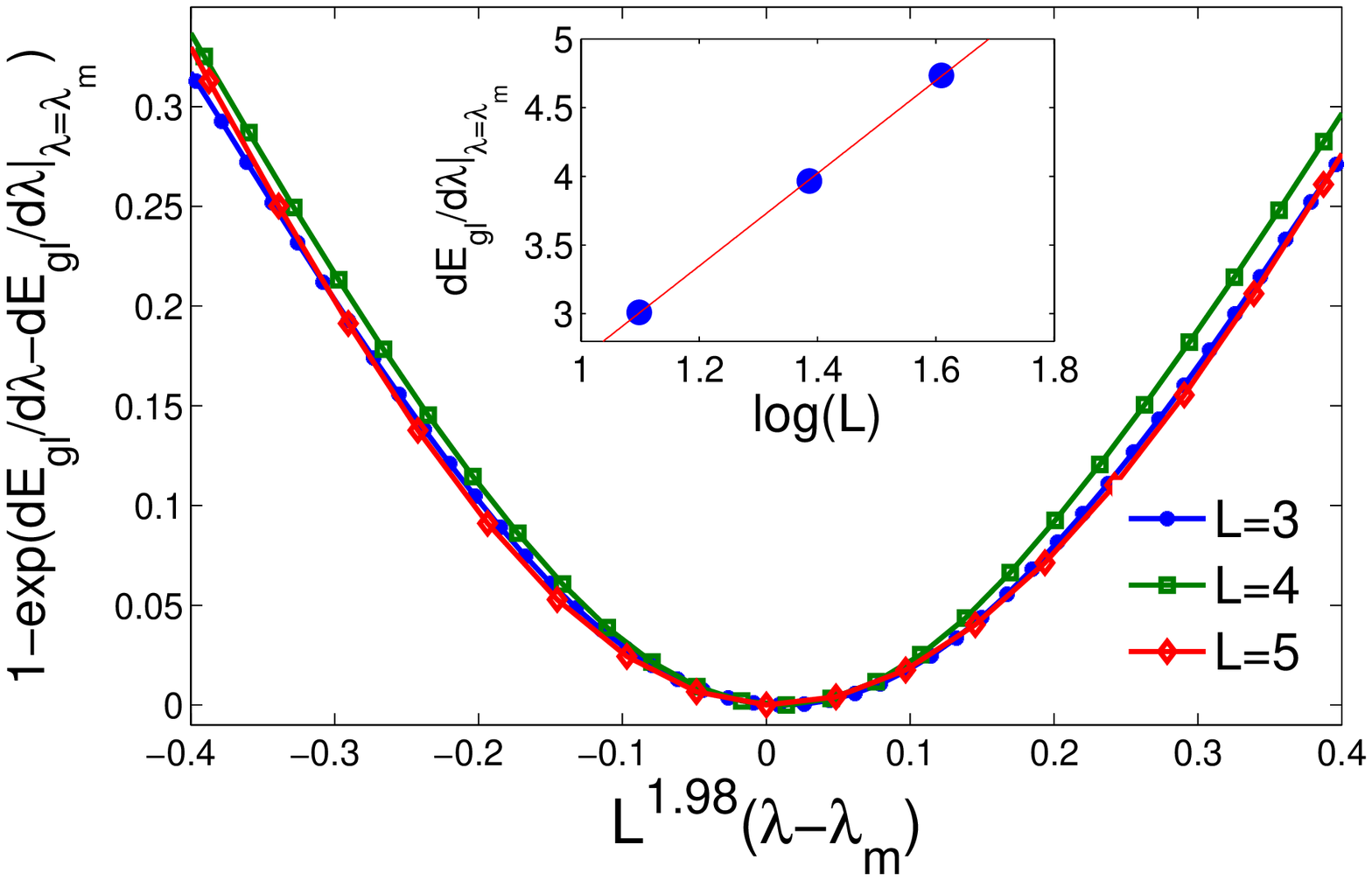}
\caption{(Color online) Finite-size scaling of global entanglement for the 2d transverse Ising model.
The inset shows the logarithmic divergence of the value of the derivative at the
maximal point $\lambda_m$.}\label{3}
\end{figure}

\begin{figure}[h]
\includegraphics[width=\columnwidth]{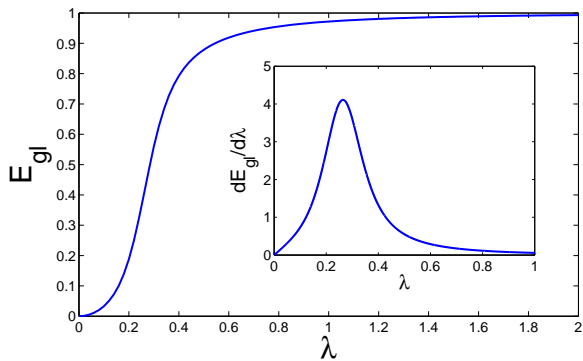}
\caption{(Color online) Global entanglement and its derivative (inset) for the $2\times2\times2$ transverse Ising system,
the only 3d system we have been able to study.}\label{3}
\end{figure}

\begin{figure}[h]
\includegraphics[width=\columnwidth]{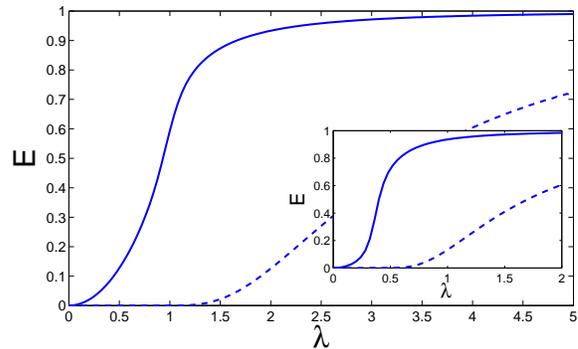}
\caption{(Color online) Global entanglement (continuous curve) and genuine entanglement (dashed curve)
for the 1d (main figure) and 2d (inset) transverse Ising model. The 1d result is for $N=16$
and the 2d result is for a $4\times4$ system.}\label{3}
\end{figure}

Finally, it is worth considering another important form of multi-partite
entanglement, namely, genuine entanglement.  Genuine entanglement in a many-particle system
represents the amount of entanglement shared by all particles.  Therefore, genuine entanglement is equal to the
$N$-tangle, the last term in Eq.~(5).  One might expect that such a term would gradually lose its
significance as $N$ increases.  However, due to the GHZ nature of our ground state, this term
($N$-tangle) is the dominant term in $E_{gl}$ and its dominance increases with increasing $\lambda$.
This is shown in Figure 8 for both the 1d and 2d models, where one can easily see that increasing
$\lambda$, increases the share of $N$-tangle in $E_{gl}$.
It is also worth noting that the structure of genuine entanglement is very similar to
an order parameter. It is zero on one side of the transition and becomes non-zero around
the critical point rising to its maximum at $1/\lambda = 0$.  This behavior becomes more pronounced
as system-sizes get larger, however, we note that the transition is not sharp and is in fact
``rounded''.  However, since net magnetization cannot be used as an order parameter here, such
behavior deserves further attention.

\section{concluding remarks}

In this paper we have studied the quantum phase transition in the
transverse field Ising model from a multipartite entanglement point of view on a one, two and
three dimensional square lattice. Our work is interesting from various points of
view. First we use a multipartite global entanglement as a measure. Secondly,
we study the symmetric GHZ-like ground state and its transition to the paramagnetic
product-state. Thirdly, by studying QPT in various dimensions we are able to establish
common features of such a transition in different universality classes.
We find that global entanglement is a good indicator of
such transitions with universal aspects, including scaling, in any dimension.
The well-suitedness of such a measure is displayed in the nice fits obtained in
figures such as Figure 2 or Figure 5, for example.
As a by-product, we find critical points
and various exponents for the 1d and 2d models consistent with
previous studies.  We note that our estimation of the critical points for the
1d and 2d models are to within one percent of the generally accepted values,
an impressive result given the limited size of the systems studied here, providing
further evidence for well-suitedness of our measure when compared with similar studies
using bi-partite measures. Our estimation of $\nu$, although acceptable, are understandably
less impressive as finite-size scaling collapses require larger system sizes
to obtain better estimates for $\nu$\cite{comm2}.  We note that
our main goal is to investigate the (universal) features of global entanglement
in quantum phase transitions, and not to produce reliable exponents for such
models.  Since the parameter $d$ determines the universality class of the systems
considered here, the fact that we see similar behavior of global entanglement at the
QPT regardless of $d$, shows what we have thus far referred to as universal features
of global entanglement.

We close by mentioning that similar studies could be carried out for
more general spin models exhibiting more complicated quantum phase transitions.
It would be interesting to see if such universal features of global entanglement carry over to other models.

\smallskip
\begin{acknowledgments}
The authors kindly acknowledge the support of Shiraz University
Research Council.  We would also like to thank A. Langari for
his comments on an earlier version of our paper.
\end{acknowledgments}


\begin{thebibliography}{99}

\bibitem{Peres} A. Peres, Found. Phys. \textbf{29}, 589 (1999).
\bibitem{Werner} R. F. Werner and M. M. Wolf, Quant. Inf. Comp. \textbf{1}, 1 (2002).
\bibitem{Plenio07} M. B. Plenio and S. Virmani, Quant. Inf. Comp. \textbf{7}, 1 (2007).
\bibitem{Ter02} B. M. Terhal, J. Theor. Comp. Science \textbf{287}, 313 (2002).
\bibitem{Van07} S. J. van Enk, N. L\"{u}tkenhaus, and H. J. Kimble, Phys. Rev. A \textbf{75}, 052318 (2007).
\bibitem{Audbook} J.~Audretsch, \emph{Entangled Systems} (Wiley, Germany, 2007).
\bibitem{Bruss02} D. Bru{\ss}, J. Math. Phys. \textbf{43}, 4237 (2002).
\bibitem{Hord07} R. Horodecki, P. Horodecki, M. Horodecki, and K. Horodecki, Rev. Mod. Phys. \textbf{81}, 865 (2009).
\bibitem{NC2000} M. A. Nielsen and I. L. Chuang, \emph{Quantum Computation and Quantum information} (Cambridge University Press, 2000).
\bibitem{Amico} L. Amico, R. Fazio, A. Osterloh, V. Vedral, Rev. Mod. Phys. \textbf{80}, 517 (2008).
\bibitem{Latorre} J. I. Latorre, A. Riera, J. Phys. A: Math. Theor. 42 (2009).
\bibitem{Arensen} M. C. Arnesen, S. Bose, and V. Vedral, Phys. Rev. Lett. \textbf{87} ,
017901 (2001).
\bibitem{Duer} W. D\"{u}r, L. Hartmann, M. Hein, M. Lewenstein, and H.
J. Briegel, Phys.Rev.Lett. \textbf{94}, 097203 (2005).
\bibitem{Guehne} O. G\"{u}hne, G. Toth, H. J. Briegel, New J. Phys. \textbf{7}, 229 (2005).
\bibitem{Wang} X.~Wang, Phys. Rev. A \textbf{64}, 012313(2001).
\bibitem{Osb02} T. J. Osborne, M. A. Nielsen, Phys.Rev. A \textbf{66}, 032110
(2002).
\bibitem{Ost02} A. Osterloh, L. Amico, G. Falci, and R. Fazio, Nature \textbf{416}, 608 (2002).
\bibitem{Vidal} G. Vidal, J. I. Latorre, E. Rico, and A. Kitaev, Phys. Rev. Lett. \textbf{90}, 227902 (2003).
\bibitem{LL06} P. Lou and J.Y. Lee, Phys. Rev. B \textbf{74}, 134402 (2006).
\bibitem{Sachdev} S. Sachdev, \emph{Quantum Phase Transitions} (Cambridge University Press, Cambridge, UK 2000).
\bibitem{GTL07} S.~Gu, G.~Tian, H.~Lin, Chin. Phys. Lett. \textbf{24} 2737 (2007).
\bibitem{OlivPRL} T. R. de Oliveira, G. Rigolin, M. C. de Oliveira, and E. Miranda Phys. Rev. Lett. \textbf{97}, 170401 (2006).
\bibitem{Wei05} T. Wei, D. Das, S. Mukhopadyay, S. Vishveshwara, and P. M. Goldbart Phys. Rev. A \textbf{71}, 060305(R) (2005).
\bibitem{Huang10} C. Huang, and F. Lin, Phy. Rev. A \textbf{81}, 032304 (2010).
\bibitem{Anfossi05} A. Anfossi, P. Giorda, A. Montorsi, and F. Traversa, Phys. Rev. Lett. \textbf{95}, 056402 (2005).
\bibitem{Oliv06} T. R. de Oliveira, G. Rigolin, and M. C. de Oliveira, Phys. Rev. A \textbf{73}, 010305(R) (2006).
\bibitem{EPR} A.~Einstein, B.~Podolsky, and N.~Rosen, Phys. Rev. \textbf{47} 777 (1935).
\bibitem{yang05} M.~Yang, Phys. Rev. A \textbf{71}, 030302(R) (2005).
\bibitem{Amico06} L.~Amico, F.~Baroni, A.~Fubini, D.~Patan, V.~Tognetti, and P.~Verruchi, Phys. Rev. A \textbf{74} 022322 (2006).
\bibitem{gold} N.~Goldenfeld, \emph{Lectures on Phase Transitions and the Renormalization Group}, (Addison Wesley, New York, 1992).
\bibitem{MW} D. A. Meyer and N. R. Wallach, J. Math. Phys. \textbf{43}, 4273 (2002).
\bibitem{Mont} A. Montakhab, and A. Asadian, Phys. Rev. A \textbf{77}, 062322 (2008).
\bibitem{langari} Lanczos diagonalization technique does not significantly improve our system-size limitations (A.~Langari, private
communications), and DMRG methods are powerful but mostly for 1d systems.
\bibitem{VidalPRL101} J.~Jordan, R.~Ortis, G.~Vidal, F. Verstraete, and J.I.~Cirac, Phys.Rev.Lett. \textbf{101}, 250602 (2008).
\bibitem{Jordan} P. Jordan and E. Wigner, Z. Phys. \textbf{47}, 631 (1928).
\bibitem{GHZ} D.M. Greenberger, M.A. Horne, and A. Zeilinger, in \emph{Bell's Theorem, Quantum Mechanics, and Conceptions
of the Universe}, (Kluwer Academics, Dordrecht, The Netherlands, 1989), pp. 73-76.
\bibitem{Buzek} P.~Stelmachovic, and V.~Buzek,  Phys. Rev. A \textbf{70}, 032313 (2004).
\bibitem{Bayat} S. Yang, A. Bayat, S. Bose, quant-ph/1005.2571.
\bibitem{sym} T.R. de Oliveira, G. Rigolin, M.C. de Oliveira, and E. Miranda, Phys. Rev. A \textbf{77}, 032325 (2008).
\bibitem{Brennen} G.K. Brennen, Quant. Inf. and Comput. \textbf{3}, 616 (2003).
\bibitem{Love} P. J. Love, A. M. van den Brink, A. Yu. Smirnov, M. H. S. Amin, M. Grajcar, E. Il'ichev, A. Izmalkov, A. M. Zagoskin, Quant. Info. Processing \textbf{6}, 187 ( 2007).
\bibitem{Boixo} S. Boixo and A. Monras, Phys. Rev. Lett. \textbf{100}, 100503 (2008).
\bibitem{Brukner} C. Brukner and A. Zeilinger, Phys. Rev. Lett. \textbf{83}, 3354 (1999).
\bibitem{Hord98} M. Horodecki and R. Horodecki, Phys.Lett. A \textbf{244}, 473
(1998); M. Horodecki, K. Horodecki, P. Horodecki, R. Horodecki, J.
Oppenheim, A. Sen(De), and U. Sen, Phys. Rev. Lett. \textbf{90},
100402 (2003); M. Horodecki, P. Horodecki, R. Horodecki, J.
Oppenheim, A. Sen(De), U. Sen, and B. Synak-Radtke, Phys. Rev. A \textbf{71}, 062307 (2005).
\bibitem{barber} M.N. Barber, in \emph{Phase Transitions and Critical Phenomena} (Academic, London, 1983), Vol.~8, pp. 146-259.
\bibitem{QMC} H.W.J.~Blote and Y.~Deng, Phys.Rev. E \textbf{66}, 066110 (2002).
\bibitem{comm2} The $d$-dimensional transverse field quantum Ising model is in the same universality class as that of the $(d+1)$ dimensional
classical Ising model[19].  Therefore, our obtained values of $\nu$ should correspond to the 2d classical Ising ($\nu=1$, exact)
and the 3d classical Ising($\nu\approx0.6$) models, respectively.

\end{thebibliography}

\end{document}